\documentstyle[12pt]{article}
\begin{document}
\thispagestyle{empty}
\begin{raggedleft}
UR-1405\\
ER-40685-852\\
hep-th/9501110\\
Jan./1995\\
\end{raggedleft}
$\phantom{x}$\vskip 0.618cm\par
{\huge \begin{center}
TWO DIMENSIONAL SUPERSYMMETRIC HARMONIC OSCILLATOR CARRYING A
REPRESENTATION OF THE GL(2$\vert$1) GRADED LIE ALGEBRA.
\end{center}}\par
\vfill
\begin{center}
$\phantom{X}$\\ {\Large A.Das\footnote{das@urhep.pas.rochester.edu} and
C.Wotzasek\footnote{clovis@urhep.pas.rochester.edu} \footnote{Permanent
address: Instituto de F\'\i sica, Universidade Federal do Rio de Janeiro,
Brasil} }\\[3ex] {\em Department of Physics and Astronomy\\ University of
Rochester\\ Rochester, NY 14627, USA}
\end{center}\par
\vfill
\begin{abstract}

\noindent We study a supersymmetric 2-dimensional harmonic oscillator
which carries a representation of the general graded Lie algebra
GL(2$\vert$1), formulate it on the  superspace, and discuss its
physical spectrum.

\end{abstract}
\vfill
\newpage

\section{Introduction}

Supersymmetry is an interesting symmetry which transforms bosons into
fermions and vice-versa \cite{susy}.  Quantum mechanical (or Classical)
theories which are supersymmetric provide realizations of graded Lie
algebras (GLA)\cite{gla}.  The most familiar GLA is the graded Poincar\'e
algebra which leads to relativistic, supersymmetric quantum field
theories, which include supergravity.

Examples of simple and extended global supersymmetries based on the
grading of space-time symmetries are abundant in simple quantum mechanical
systems - the one dimensional supersymmetric harmonic oscillator being the
simplest example of such systems\cite{ssqm}\cite{ssqm2}.  There exist,
however, many other GLA's  which involve grading internal symmetry
algebras.  The most familiar of such GLA's are the  OSp(2m$\vert$n) and
SL(m$\vert$n)\cite{intmod}.  While realizations of such algebras arise
naturally in integrable models, there does not yet exist a quantum or
classical mechanical realization of the most general graded Lie algebra,
namely, GL(m$\vert$n).  In this paper, we construct a supersymmetric two
dimensional harmonic oscillator which provides a realization of
GL(2$\vert$1) as its symmetry algebra.  In section II, we discuss the
GL(2$\vert$1) algebra with raising and lowering operators.  In section III
we present our model of a supersymmetric harmonic oscillator and discuss
all the symmetries associated with this system.  We show that the symmetry
algebra coincides with GL(2$\vert$1).  In section IV, we discuss the
spectrum of states associated with this Hamiltonian and present a
superspace description of this theory.  Finally, we discuss our
conclusions in section V.

\section{Graded Lie Algebra GL(2$\vert$1)}

Graded Lie algebras\cite{gla} include both bosonic and fermionic generators
satisfying commutation and anticommutation relations respectively and have
the following general structure:

\begin{eqnarray}
\label{gla}
\left[ B_m , B_n \right]_-& = & f_{mn}^k B_k\nonumber\\
\left[ B_m \;, F_\alpha \right]_-& = & h_{m \alpha}^{\beta} F_\beta\\
\left[F_\alpha \;\;, F_\beta \right]_{+}& = & g_{\alpha \beta}^m B_m\nonumber
\end{eqnarray}

\noindent with the brackets $\left[\dots , \dots\right]_{\mp}$ denoting
commutators and anti-commutators respectively, $k,\:m,\:n=1,\:2,
\ldots,\:N$, and $\alpha,\:\beta=1,\:2,\dots,\:M$.  The even or bosonic
generators $B_m$ form the underlying
Lie algebra, while the odd or fermionic generators $F_\alpha$ provide a
grading of this algebra consistent with the generalized Jacobi identities.

In this section we shall study the graded Lie algebra GL(2$\vert$1) whose
underlying bosonic algebra is GL(2)$\oplus$GL(1).  Here we shall use the
boson/fermion
representation obtained with help of canonical realizations, i.e.,
realizations in terms of pairs of boson/fermion creation and annihilation
operators satisfying canonical (anti)commutation relations.  Consider
the set of bosonic and fermionic operators

\begin{equation}
\label{cao}
\{a_k^{\dagger} , a_k \; ; \; k=1,2 \} \;\;\mbox{and}\;\;\{ a_3^{\dagger}
, a_3\}
\end{equation}

\noindent satisfying the canonical (anti)commutation relations

\begin{eqnarray}
\label{ccr}
\left[a_k,a_m^{\dagger}\right]_{-} &=& \delta_{km} \nonumber\\
\left[a_3,a_3^{\dagger}\right]_{+} &=& 1
\end{eqnarray}

\noindent with all other (anti)commutators vanishing.  The four bilinear
operators $B \sim a_k^{\dagger}a_m$ define the generators of the
underlying GL(2) algebra,which together with  $B \sim a_3^{\dagger}a_3$
constitute the five bosonic generators of the GL(2$\vert$1) algebra.  The
four fermionic generators of this algebra are defined by the bilinear
operators $F \sim a_3^{\dagger}a_k$ and $F \sim a_k^{\dagger}a_3$.  All
together these nine operators generate the $Z_2$ graded GL(2$\vert$1)
algebra.  It is a simple task to verify that they satisfy the algebra
(\ref{gla}).  For instance, if we denote these nine operators as

(a) Bosonic Generators

\begin{eqnarray}
\label{bosgen}
B_1 &=& a_1^{\dagger}a_1\nonumber\\
B_2 &=& a_1^{\dagger}a_2\nonumber\\
B_3 &=& a_2^{\dagger}a_1\nonumber\\
B_4 &=& a_2^{\dagger}a_2\nonumber\\
B_5 &=& a_3^{\dagger}a_3
\end{eqnarray}

(b) Fermionic Generators

\begin{eqnarray}
\label{fergen}
F_1 &=& a_1^{\dagger}a_3\nonumber\\
F_2 &=& a_2^{\dagger}a_3\nonumber\\
F_3 &=& a_3^{\dagger}a_1\nonumber\\
F_4 &=& a_3^{\dagger}a_2
\end{eqnarray}

\noindent it is then a simple task to find the nonvanshing
structure constants in (\ref{gla}).

For future convenience, we introduce a new basis of the fermionic
generators as

\begin{eqnarray}
\label{oddcharges}
Q_{\scriptscriptstyle R}&=& {1\over{\sqrt2}}\left(a_1^{\dagger}a_3 - i
a_2^{\dagger} a_3\right)\nonumber\\
\overline Q_{\scriptscriptstyle R} &=& {1\over{\sqrt2}}
\left(a_3^{\dagger}a_1 + i a_3^{\dagger}a_2\right)
\nonumber\\
Q_{\scriptscriptstyle L} &=& {i\over{\sqrt2}}\left(a_3^{\dagger}a_1 - i
a_3^{\dagger}a_2\right)
\nonumber\\
\overline Q_{\scriptscriptstyle L} &=&
{{-i}\over{\sqrt2}}\left(a_1^{\dagger}a_3 + i a_2^{\dagger} a_3
\right)
\end{eqnarray}

\noindent The anti-commutation relations amongst these charges are
easily computed.  We also redefine the five bosonic operators as

\begin{eqnarray}
\label{evencharges}
h_b &=&   a_1^{\dagger} a_1 + a_2^{\dagger}a_2 \nonumber\\
h_f &=&    a_3^{\dagger} a_3 \nonumber\\
\Delta_1 &=&   a_1^{\dagger}a_1 - a_2^{\dagger}a_2 \nonumber\\
i\;\Delta_2 &=&   a_2^{\dagger}a_1 - a_1^{\dagger}a_2 \nonumber\\
\Delta_3 &=&   a_1^{\dagger}a_2 + a_2^{\dagger} a_1
\end{eqnarray}

\noindent and introduce the operator $H = h_b + h_f $.  The algebra
of the fermionic charges (\ref{oddcharges}) becomes

\begin{eqnarray}
\label{oddalgebra2}
\left[Q_{\scriptscriptstyle R}, \overline Q_{\scriptscriptstyle
R}\right]_+ &=& {1\over2} \left( H + \Delta_2 + h_f\right)
\nonumber\\
\left[Q_{\scriptscriptstyle L}, \overline Q_{\scriptscriptstyle
L}\right]_+ &=& {1\over2}
\left( H - \Delta_2 + h_f\right)\nonumber\\
\left[Q_{\scriptscriptstyle R}, Q_{\scriptscriptstyle L}\right]_+ &=&
{1\over2} \left(\Delta_3 + i\;\Delta_1
\right)\nonumber\\
\left[\overline Q_{\scriptscriptstyle R}, \overline Q_{\scriptscriptstyle
L}\right]_+ &=& {1\over2} \left(
\Delta_3 - i\;\Delta_1\right)
\end{eqnarray}

Similarly, the algebra of the new bosonic charges (\ref{evencharges}) can be
computed straightforwardly, to give

\begin{eqnarray}
\label{bosonalgebra}
\left[\Delta_k,\Delta_m\right]_- &=& 2\; i \; \epsilon_{kmn}\; \Delta_n
\nonumber\\
\left[h_b,\Delta_m\right]_- &=& 0\nonumber\\
\left[h_f,\Delta_m\right]_- &=& 0\nonumber\\
\left[h_b, h_f\right]_- &=& 0
\end{eqnarray}

\noindent while the remaining nonvanishing boson-fermion commutation
relations are

\begin{eqnarray}
\label{boson-fermion1}
\left[h_f,Q_{\scriptscriptstyle R}\right]_- &=& -\;
\left[h_b,Q_{\scriptscriptstyle R}\right]_- = -\;Q_{\scriptscriptstyle
R}\nonumber\\
\left[h_f,\overline Q_{\scriptscriptstyle R}\right]_- &=& -
\;\left[h_b,\overline Q_{\scriptscriptstyle R}\right]_- =+\;
\overline Q_{\scriptscriptstyle R} \nonumber\\
\left[h_f,Q_{\scriptscriptstyle L}\right]_- &=& -\;
\left[h_b,Q_{\scriptscriptstyle L}\right]_- =+\; Q_{\scriptscriptstyle
L}\nonumber\\
\left[h_f,\overline Q_{\scriptscriptstyle L}\right]_- &=& -\;
\left[h_b,\overline Q_{\scriptscriptstyle L}\right]_- = -\;\overline
Q_{\scriptscriptstyle L}
\end{eqnarray}

\begin{eqnarray}
\label{boson-fermion2}
\left[\Delta_1,Q_{\scriptscriptstyle R}\right]_- &=& +i\;\overline
Q_{\scriptscriptstyle L}\nonumber\\
\left[\Delta_1,\overline Q_{\scriptscriptstyle R}\right]_- &=&
+i\;Q_{\scriptscriptstyle L}\nonumber\\
\left[\Delta_1,Q_{\scriptscriptstyle L}\right]_- &=& -i\;{\overline
Q_{\scriptscriptstyle R}}\nonumber\\
\left[\Delta_1,\overline Q_{\scriptscriptstyle L}\right]_- &=& -i\;
Q_{\scriptscriptstyle R}
\end{eqnarray}

\begin{eqnarray}
\label{boson-fermion3}
\left[\Delta_2,Q_{\scriptscriptstyle R}\right]_- &=& +\;
Q_{\scriptscriptstyle R}\nonumber\\
\left[\Delta_2,\overline Q_{\scriptscriptstyle R}\right]_- &=&
-\;\overline Q_{\scriptscriptstyle R}\nonumber\\
\left[\Delta_2,Q_{\scriptscriptstyle L}\right]_- &=&+\;
Q_{\scriptscriptstyle L}\nonumber\\
\left[\Delta_2,\overline Q_{\scriptscriptstyle L}\right]_- &=&
-\;\overline Q_{\scriptscriptstyle L}
\end{eqnarray}

\begin{eqnarray}
\label{boson-fermion4}
\left[\Delta_3,Q_{\scriptscriptstyle R}\right]_- &=& +\;\overline
Q_{\scriptscriptstyle L}\nonumber\\
\left[\Delta_3,\overline Q_{\scriptscriptstyle R}\right]_- &=&
-\;Q_{\scriptscriptstyle L}\nonumber\\
\left[\Delta_3,Q_{\scriptscriptstyle L}\right]_- &=& -\;{\overline
Q_{\scriptscriptstyle R}}\nonumber\\
\left[\Delta_3,\overline Q_{\scriptscriptstyle L}\right]_- &=& +\;
Q_{\scriptscriptstyle R}
\end{eqnarray}

\noindent Using these relations we can verify that all generators satisfy
the generalized Jacobi identities.  Observe from eqs.(\ref{bosonalgebra})
and (\ref{boson-fermion1})
that all nine generators of the GL(2$\vert$1) graded Lie algebra described
above commute with $H$ which stays in the center of the algebra.

\section{Two Dimensional Supersymmetric Harmonic Oscillator}

In this section we introduce our model, a two-dimensional supersymmetric
harmonic oscillator which, as mentioned in the introduction, carries a
representation of the graded GL(2$\vert$1) algebra described in section II.
This model is described by the following Lagrangian

\begin{equation}
\label{lagrangian}
L = {1 \over 2}\left(\dot q^T \dot q - q^T q\right) + {i\over 2}
\psi^T \left({d\over{dt}}- i\;\sigma_2\right) \psi
\end{equation}

\noindent where

\begin{equation}
\label{boscoord}
q=\left(
\begin{array}{c}
q_1\\
q_2
\end{array}
\right)
\end{equation}

\noindent and

\begin{equation}
\label{fercoord}
\psi=\left(
\begin{array}{c}
\psi_1\\
\psi_2
\end{array}
\right)
\end{equation}

\noindent are the oscillator's bosonic and fermionic coordinates, which
are assumed to be real.  We have chosen the mass and the frequency to be
unity, for simplicity.  Here $\sigma_k$ stands for the Pauli matrices and
$q^T$ and $\psi^T$ stand for matrix transposition, as usual.  Note that up
to total derivatives, we can also write the Lagrangian (\ref{lagrangian})
as

\begin{equation}
\label{lagrangian2}
L = - {1\over 2} q^T \left({d\over{dt}} - i\; \sigma_2 \right)
\left({d\over{dt}} + i\; \sigma_2 \right)q +{i\over 2} \psi^T
\left({d\over{dt}} - i\; \sigma_2 \right)\psi
\end{equation}

\noindent It is important to mention that compared to the {\it usual} two
dimensional supersymmetric harmonic oscillator, the model under
investigation here is constructed with half the number of fermionic
degrees of freedom, i.e, it has two second-order bosonic variables (or
four first-order) and two first-order fermionic variables.  In the
two-dimensional matrix space of Eqs.(\ref{boscoord},\ref{fercoord}), let
us denote a complete basis of real, (2x2) matrices by

\begin{equation}
\label{basis}
\tau_a =\left(\sigma_0,\sigma_1,i\;\sigma_2,\sigma_3\right)\;;\;\;a=0,1,2,3
\end{equation}

\noindent where  $\sigma_0$ is the (2x2) identity matrix. It is
straightforward to show that the action of the theory described by
(\ref{lagrangian}) is invariant under the four supersymmetry
transformations

\begin{eqnarray}
\label{susy1}
\delta q &=& \varepsilon _a \tau_a \psi \nonumber\\
\delta \psi &=& i\left({d\over{dt}} + i\;
\sigma_2\right)\varepsilon_a\tau_a^T q
\end{eqnarray}

\noindent with $\varepsilon_a$ being four infinitesimal, constant
Grassmann parameters that characterize the transformations.  In the
Hamiltonian language\cite{das}, which is more appropriate for our
purposes, the Hamiltonian operator is given by

\begin{equation}
\label{hamiltonian}
H={1\over2}\left(p^Tp+q^Tq\right) -{1\over2}\psi^T\sigma_2\psi
\end{equation}

\noindent and enjoys the following set of global invariances

\noindent(A) Supersymmetry:

\begin{eqnarray}
\label{susy2}
\delta q &=& {1\over\sqrt2} \varepsilon_a \tau_a \psi\nonumber\\
\delta \psi &=&  {i\over\sqrt2}\varepsilon_a\left(\tau_a^T p + i\;
\sigma_2\tau_a^T q\right)\nonumber\\
\delta p &=& {{i}\over\sqrt2}\varepsilon_a\tau_a\sigma_2\psi
\end{eqnarray}

\noindent The Noether supercharges generating these transformations are
given by

\begin{equation}
\label{susycharge}
Q_a = {1\over\sqrt2}\left(p^T\tau_a \psi - i\; q^T\tau_a\sigma_2\psi\right)
\end{equation}

\noindent Indeed, note that given the generalized Dirac brackets (see
Eq.(\ref{dirac-brack}))

\begin{eqnarray}
\label{ccr2}
\left\{q_k,p_m\right\} &=& \delta_{km}\nonumber\\
\left\{\psi_\alpha ,\psi_\beta\right\} &=& -i\:\delta_{\alpha\beta}
\end{eqnarray}

\noindent we obtain the supersymmetry transformations (\ref{susy2})
above as

\begin{equation}
\label{variation}
\delta A = \left\{A,\varepsilon_aQ_a\right\}
\end{equation}

\noindent where $A$ stands for $q_k$, $p_k$ and $\psi_\alpha$.  The
invariance of the Hamiltonian implies that $\left\{H, Q_a\right\}=0$ which
in turn shows that the $Q_a$'s are constants of motion.  Besides the four
supersymmetries above, this Hamiltonian is also invariant under the
following global symmetries.

\noindent (B) Rotation on the four-dimensional (bosonic) phase-space
plane:

(i)  The transformations

\begin{eqnarray}
\label{rotation}
\delta q &=& - i\;\alpha \sigma_2 q \nonumber\\
\delta p &=& - i\;\beta\sigma_2 p
\end{eqnarray}

\noindent with $\alpha$,$\beta$ bosonic, constant infinitesimal parameters
clearly are a symmetry of $H$.  However, in order to preserve the canonical
commutation relations (\ref{ccr2}) we must have $\alpha=\beta$.  These
transformations are generated by the angular momentum operator

\begin{equation}
\label{angmom}
J_1 = {i\over 2}\left(q^T\sigma_2p-p^T\sigma_2q\right)
\end{equation}

\noindent Note that these transformations do not mix coordinate and
momentum variables.

(ii)  The transformations

\begin{eqnarray}
\label{mix}
\delta q &=& \lambda_a \tau_a p\nonumber\\
\delta p &=& - \lambda_a \tau_a q
\end{eqnarray}

\noindent with $\lambda_a$ constant, bosonic, infinitesimal parameters are
also a set of symmetries of the Hamiltonian which preserve the Dirac
brackets relations, as long as $a\neq 2$.  The charges generating these
transformations are

\begin{eqnarray}
\label{mix-charges}
L_0 &=& {1\over2}\left(p^Tp + q^Tq\right)\nonumber\\
L_1 &=& {1\over2}\left(p^T\sigma_1 p + q^T\sigma_1 q\right)\nonumber\\
L_3 &=& {1\over2}\left(p^T\sigma_3 p + q^T\sigma_3 q\right)
\end{eqnarray}

\noindent (C)  Rotation on the two-dimensional fermionic phase-space.

The generalized rotation in the fermionic phase-space

\begin{equation}
\label{ferrotation}
\delta \psi = \vec{ \alpha}\cdot\vec{\sigma}\;\psi
\end{equation}

\noindent with $\vec{ \alpha}=\left(\alpha_1,\alpha_2,\alpha_3\right)$
being a set of constant bosonic parameters, is another group of symmetries
in the action.  Again, preservation of the canonical commutation
relations (\ref{ccr2}) imposes $\alpha_1=\alpha_3=0$ as conditions over
the possible values that these parameters can take.  These
transformations are generated by the following charge,

\begin{equation}
\label{ferangmom}
J_2={1\over2}\psi^T\sigma_2\psi
\end{equation}

\noindent representing the fermionic contribution to the total angular
momentum.

Now, in order to make contact with the graded algebra GL(2$\vert$1)
defined in the
preceeding section, let us redefine these charges as

\begin{eqnarray}
\label{newbos}
h_b &=& L_0 = {1\over2}\left(p^Tp + q^Tq\right)\nonumber\\
h_f &=& - J_2 = -{1\over2}\psi^T\sigma_2\psi\nonumber\\
\Delta_1 &=& L_1 = {1\over2}\left(p^T\sigma_1 p + q^T\sigma_1 q\right)
\nonumber\\
\Delta_2 &=& J_1 = {i\over
2}\left(q^T\sigma_2p-p^T\sigma_2q\right)\nonumber\\
\Delta_3 &=&  L_3 = {1\over2}\left(p^T\sigma_3 p + q^T\sigma_3 q\right)
\end{eqnarray}

\noindent The canonical Dirac bracket algebra of these charges read

\begin{eqnarray}
\label{nbosalg}
\left\{\Delta_k,\Delta_m\right\} &=& 2\;\epsilon_{kmn}\Delta_n\nonumber\\
\left\{h_b,\Delta_k\right\} &=& 0\nonumber\\
\left\{h_f,\Delta_k\right\} &=& 0
\end{eqnarray}

\noindent and can be seen to agree with the bosonic sector of the
GL(2$\vert$1) algebra, eq.(\ref{bosonalgebra}), when the classical Dirac
bracket algebra is quantized through the usual replacement
$\left\{\ldots,\ldots\right\}\rightarrow
-i\:\left[\ldots,\ldots\right]_{\mp}$.

Next we redefine the supersymmetry generators (\ref{susycharge}) as

\begin{eqnarray}
\label{holom}
Q_{\scriptscriptstyle R} &=& {1\over2} \left(Q_0 +i\;Q_2\right)\nonumber\\
&=& {1\over\sqrt2}\left(p^T+i\;q^T\right)T_+\psi\nonumber\\
\overline Q_{\scriptscriptstyle R} &=& {1\over2} \left(Q_0
-i\;Q_2\right)\nonumber\\ &=&
{1\over\sqrt2}\left(p^T-i\;q^T\right)T_-\psi\nonumber\\
Q_{\scriptscriptstyle L} &=& {1\over2} \left(Q_1 +i\;Q_3\right)\nonumber\\
&=& {1\over\sqrt2}\left(p^T-i\;q^T\right)\sigma_1 T_+\psi\nonumber\\
\overline Q_{\scriptscriptstyle L} &=& {1\over2} \left(Q_1
-i\;Q_3\right)\nonumber\\ &=&
{1\over\sqrt2}\left(p^T+i\;q^T\right)\sigma_1 T_-\psi\nonumber\\
\end{eqnarray}

\noindent where $T_{\pm}={1\over2}\left(1\mp \sigma_2\right)$ is the
 ``helicity'' projection operator.  Using the genralized Dirac brackets,
Eq.(\ref{ccr2}), we can verify that the nine symmetry generating operators
in (\ref{newbos}) and (\ref{holom}), possess an algebra whose quantum
version is isomorphic to that presented in section II.  Moreover, if we
introduce, as usual, the representation of the phase-space variables $q_k$
and $p_k$ in terms of creation and annihilation operators as

\begin{eqnarray}
\label{cao2}
a_k &=& {1\over\sqrt2}\left(q_k +i\;p_k\right)\nonumber\\
a_k^{\dagger} &=& {1\over\sqrt2}\left(q_k - i\; p_k\right)
\end{eqnarray}

\noindent and define

\begin{eqnarray}
\label{fcao}
a_3 &=& {i\over\sqrt2}\left(\psi_1 +i\;\psi_2\right)\nonumber^\\
a_3^{\dagger} &=& {{-i}\over\sqrt2}\left(\psi_1 - i\;\psi_2\right)
\end{eqnarray}

\noindent which, by virtue of (\ref{ccr2}) do satisfy (\ref{ccr}), then we
can write the generators of symmetry in (\ref{newbos}) and
(\ref{holom}) in the same form as the graded Lie algebra generators
(\ref{oddcharges}) and (\ref{evencharges}), defined in section II.
It becomes a matter of simple calculation to verify that these nine
charges (five bosonic and four fermionic) satisfy exactly the graded Lie
Algebra GL(2$\vert$1) found in section II.

\section{Spectrum And Superspace Formulation}

Let us examine in this section the action of the GL(2$\vert$1) operators,
defined in the previous sections, on the states of the Hilbert space of
the quantum mechanical model.  The spectrum of the normal ordered theory is
given by $\{ { \cal E}_n,\mid n> \}$, where the eigenvalues and
eigenvectors are

\begin{eqnarray}
\label{spectrum}
{\cal E}_n &=& n_+ + n_- + n_f\nonumber\\
\mid n > &=& \mid n_+ , n_- , n_f >
\end{eqnarray}

\noindent with $n_{\pm}=1,2,\ldots\;$ and $n_f=0,1$.  Conventionally, the
states with $n_f=0$(1) are called bosonic (fermionic).  Here
$n_{\pm}\mbox{ and } n_f$ are the eigenvalues of the bosonic and
fermionic number operators, $N_{\pm}=a^{\dagger}_{\pm}a_{\pm}$ and
$N_f=a^{\dagger}_3a_3$, and

\begin{equation}
\label{chiral}
a_{\pm}={1\over \sqrt 2}\left(a_1 \pm i\:a_2\right)
\end{equation}

\noindent  By inspection we see that the ground state is a non-degenerate
bosonic state with zero energy.
The first excited level is three fold degenerate, possessing one fermionic
and two bosonic states.  The second excited energy level is five fold
degenerate, with two fermionic and three bosonic states, and so on.
The states of the first few levels are displayed below:

\begin{eqnarray}
\label{states}
\mid 0> &=& \mid 0,0,0>\nonumber\\
\mid 1> &=& \{\mid 1,0,0>\:,\:\mid 0,0,1>\:,\:\mid 0,1,0>\}\nonumber\\
\mid 2> &=& \{\mid 2,0,0>\:,\:\mid 1,0,1>\:,\:\mid 1,1,0>\:,\:\mid 0,1,1>
\:,\:\mid 0,2,0>\}\\
&\vdots&\nonumber
\end{eqnarray}

\noindent  In terms of the (chiral) operators (\ref{chiral}), the
GL(2$\vert$1) generators (\ref{oddcharges}) and (\ref{evencharges}) read

\begin{eqnarray}
\label{gl21-chiral}
Q_{\scriptscriptstyle R} &=& a_+^{\dagger} a_3\nonumber\\
\overline Q_{\scriptscriptstyle R} &=& a_+ a_3^{\dagger}\nonumber\\
Q_{\scriptscriptstyle L} &=& i \; a_- a_3^{\dagger}\nonumber\\
\overline Q_{\scriptscriptstyle L} &=& -i \; a_-^{\dagger} a_3\nonumber\\
\Delta_1 &=& a_+^{\dagger}a_- + a_-^{\dagger}a_+\nonumber\\
\Delta_2 &=& a_+^{\dagger}a_+ - a_-^{\dagger}a_-\nonumber\\
i\;\Delta_3 &=& a_-^{\dagger}a_+ - a_+^{\dagger}a_-\nonumber\\
h_b &=& a_+^{\dagger}a_+ + a_-^{\dagger}a_-\nonumber\\
h_f &=& a_3^{\dagger}a_3
\end{eqnarray}

\noindent  On the degenerate levels the supersymmetry
generators take bosonic states into fermionic ones and vice versa as

\begin{eqnarray}
Q_{\scriptscriptstyle R} \mid n_+,n_-, n_f> & = & \sqrt{n_+ + 1}\;
\delta_{n_f,1}\mid n_+ +1, n_-, n_f -1>\nonumber\\
\overline Q_{\scriptscriptstyle R} \mid n_+,n_-, n_f> & = & \sqrt{n_+}\;
\delta_{n_f,0}\mid n_+ -1,n_-, n_f +1>\nonumber\\
Q_{\scriptscriptstyle L} \mid n_+,n_-, n_f> &=& \sqrt{n_-}\;\delta_{n_f,0}
\mid n_+ , n_- -1, n_f +1 >\nonumber\\
\overline Q_{\scriptscriptstyle L} \mid n_+,n_-, n_f> &= & \sqrt{n_- + 1}\;
\delta_{n_f,1}\mid n_+ ,n_- +1, n_f -1>
\end{eqnarray}

\noindent Quite clearly the supersymmetry charge $Q_{\scriptscriptstyle R}$
($\overline Q_{\scriptscriptstyle R}$)
creates
(destroys) a right-handed boson and destroys (creates) a fermion.
Similarly the $Q_{\scriptscriptstyle L}$ ($\overline Q_{\scriptscriptstyle L}$)
destroys (creates) a left handed
boson while creating (destroying) a fermion.  Note that
$Q_{\scriptscriptstyle R}$ and $\overline
Q_{\scriptscriptstyle R}$ ($Q_{\scriptscriptstyle L}$ and $\overline
Q_{\scriptscriptstyle L}$) have no effect on the $n_-$ ($n_+$)
eigenvalues, showing the existence of a chiral supersymmetry which,
ultimately, is due to the fact that we have twice as many bosonic
variables as the fermionic ones.  The supersymmetry charges can be used to
generate the states in a given level once the highest state is given.
Then, starting from the state $\mid 0,n,0>$, one can generate all the
states belonging to the ${\cal E}_n$ subspace by consecutive applications
of $Q_{\scriptscriptstyle L}$ and $Q_{\scriptscriptstyle L}$ until the
state $\mid n,0,0>$ is reached.

\begin{equation}
\mid n,0,0>\buildrel{Q_{\scriptscriptstyle R}}\over\longleftarrow \mid n-1,0,1>
\buildrel{Q_{\scriptscriptstyle L}}\over\longleftarrow\mid n-1,1,0> \dots\mid
0,n-1,1>
\buildrel{Q_{\scriptscriptstyle L}}\over\longleftarrow\mid 0,n,0>
\end{equation}

\noindent  Similarly, starting with the state $\mid
n,0,0>$ and using consecutively the charges $\overline
Q_{\scriptscriptstyle R}$ and $\overline Q_{\scriptscriptstyle L}$ one
generates the whole subspace ${\cal E}_n$ again.

\begin{equation}
\mid n,0,0>\buildrel{\overline Q_{\scriptscriptstyle R}}
\over\longrightarrow \mid n-1,0,1>
\buildrel{\overline Q_{\scriptscriptstyle L}}\over\longrightarrow\mid n-1,1,0>
\dots\mid 0,n-1,1>
\buildrel{\overline Q_{\scriptscriptstyle L}}\over\longrightarrow\mid 0,n,0>
\end{equation}

\noindent  This action of the supersymmetry charges is easily seem on the set
of states shown in equation
(\ref{states}).

The action of the
bosonic operators on the other hand only connect states with the same
fermion number.  The operators $h_f$, $h_b$, and $\Delta_2$ are diagonal
in the chiral basis (\ref{spectrum}):

\begin{eqnarray}
\label{diagonal}
h_b \mid n_+,n_-,n_f> &=& (n_+ + n_-)\mid n_+,n_-,n_f>\nonumber\\
\Delta_2 \mid n_+,n_-,n_f> &=& (n_+ -n_-)\mid n_+,n_-,n_f>\nonumber\\
h_f \mid n_+,n_-,n_f> &=& n_f \mid n_+,n_-,n_f>
\end{eqnarray}

\noindent  These operators have the usual interpretation as bosonic and
fermionic Hamiltonians ($h_b$ and $h_f$), and chirality operator
($\Delta_2$).  Finally, the non-diagonal operators $\Delta_1$ and $\Delta_3$
being bosonic in nature only connect states of the same Grassman parity but
with chiralities two unities different from the initial state:

\begin{eqnarray}
\label{non-diag}
{1\over 2}(\Delta_1 +i\Delta_3)\mid n_+,n_-,n_f> &=& \sqrt{(n_-+1)n_+}
\mid n_+ -1, n_-+1,n_f>\nonumber\\
{1\over 2}(\Delta_1 -i\Delta_3) \mid n_+,n_-,n_f> &=&
\sqrt{(n_+ +1)n_-}\mid n_++1,n_--1,n_f>
\end{eqnarray}

We finish this section with a discussion of the superspace formulation of
this problem.  To this end we rewrite the supersymmetry transformations
(\ref{susy1}) in
terms of the transformations generated by the chiral supersymmetry
charges (\ref{holom}), which seem to be more appropriate for this model.
The transformations generated by $Q_{\scriptscriptstyle R}$, $\overline
Q_{\scriptscriptstyle R}$, $Q_{\scriptscriptstyle L}$ and
$\overline Q_{\scriptscriptstyle L}$ are, respectively

\begin{eqnarray}
\label{r-chiral-transf}
\delta_{\scriptscriptstyle R}\: q &=& \varepsilon_{\scriptscriptstyle R}\:
T_+ \:\psi\nonumber\\
\delta_{\scriptscriptstyle R}\: \psi &=& \varepsilon_{\scriptscriptstyle R}\:
T_- \:{\cal D}_+ q\\
\mbox{}\nonumber\\
\overline\delta_{\scriptscriptstyle R}\: q &=&
\overline\varepsilon_{\scriptscriptstyle R}\: T_- \:\psi\nonumber\\
\overline\delta_{\scriptscriptstyle R}\:\psi &=&\overline
\varepsilon_{\scriptscriptstyle R}\: T_+\: {\cal D}_- q
\end{eqnarray}

\begin{eqnarray}
\label{l-chiral-transf}
\delta_{\scriptscriptstyle L}\: q &=& \varepsilon_{\scriptscriptstyle L}\:
T_+\: \sigma_1\psi\nonumber\\
\delta_{\scriptscriptstyle L}\: \psi &=& \varepsilon_{\scriptscriptstyle L}\:
T_+\:\sigma_1 {\cal D}_- q\\
\mbox{}\nonumber\\
\overline\delta_{\scriptscriptstyle L}\: q &=&
\overline\varepsilon_{\scriptscriptstyle L}\: T_-\: \sigma_1\psi\nonumber\\
\overline\delta_{\scriptscriptstyle R}\: \psi &=&
\overline \varepsilon_{\scriptscriptstyle L}\: T_-\: \sigma_1 {\cal D}_+ q
\end{eqnarray}

\noindent Here we have introduced the notation ${\cal D}_{\pm}=
i(\partial_t \pm i)$.  To obtain these four supersymmetries in a
superfield language, we introduce two Grassman variables for each chiral
sector as $\theta_{\scriptscriptstyle R}$, $\overline
\theta_{\scriptscriptstyle R}$, $\theta_{\scriptscriptstyle L}$ and
$\overline \theta_{\scriptscriptstyle L}$, and define two chiral
superfields $\phi_{\scriptscriptstyle R}$ and $\phi_{\scriptscriptstyle
L}$.  The transformation in the right chiral sector can be obtained from
the following superfield and (differential operator) supercharge

\begin{eqnarray}
\label{r-superfield}
\phi_{\scriptscriptstyle R} &=& q + \theta_{\scriptscriptstyle R} T_+ \psi
+\overline\theta_{\scriptscriptstyle R} T_- \psi\nonumber\\
Q_{\scriptscriptstyle R} &=& T_+
\frac{\partial}{\partial\theta_{\scriptscriptstyle
R}}-\overline\theta_{\scriptscriptstyle R} T_- {\cal D}_+\nonumber\\
\overline Q_{\scriptscriptstyle R} &=& T_-
\frac{\partial}{\partial\overline\theta_{\scriptscriptstyle
R}}-\theta_{\scriptscriptstyle R} T_+ {\cal D}_-
\end{eqnarray}

\noindent while those in the left chiral sector come from

\begin{eqnarray}
\label{l-superfield}
\phi_{\scriptscriptstyle L} &=& q + \theta_{\scriptscriptstyle L}
T_+\sigma_1 \psi +\overline\theta_{\scriptscriptstyle L} T_- \sigma_1
\psi\nonumber\\
Q_{\scriptscriptstyle L} &=& T_+
\frac{\partial}{\partial\theta_{\scriptscriptstyle
L}}-\overline\theta_{\scriptscriptstyle L} T_- {\cal D}_-\nonumber\\
\overline Q_{\scriptscriptstyle L} &=& T_-
\frac{\partial}{\partial\overline\theta_{\scriptscriptstyle
L}}-\theta_{\scriptscriptstyle L}T_+ {\cal D}_+
\end{eqnarray}

\noindent These transformations can be organized in a matrix like
structure with the following form

\begin{eqnarray}
\label{matrix-transf}
\delta \Phi &=& {\bf \epsilon}^T {\bf Q} \Phi\nonumber\\
\overline\delta \Phi &=& {\bf \overline\epsilon}^T {\bf \overline Q}
\Phi
\end{eqnarray}

\noindent where

\begin{equation}
\label{q-matrix}
{\bf Q}=\left(
\begin{array}{cc}
Q_{\scriptscriptstyle R} & 0 \\
0 & Q_{\scriptscriptstyle L}
\end{array}
\right)
\end{equation}

\begin{equation}
\label{qbar-matrix}
{\bf\overline Q}=\left(
\begin{array}{cc}
\overline Q_{\scriptscriptstyle R} & 0 \\
0 & \overline Q_{\scriptscriptstyle L}
\end{array}
\right)
\end{equation}

\noindent are block-diagonal (4x4) matrices and

\begin{equation}
\label{col-field}
\Phi=\left(
\begin{array}{c}
\phi_{\scriptscriptstyle R}\\
\phi_{\scriptscriptstyle L}
\end{array}
\right)
\end{equation}

\begin{equation}
\label{par-field}
{\bf\epsilon}=\left(
\begin{array}{c}
\varepsilon_{\scriptscriptstyle R}\\
\varepsilon_{\scriptscriptstyle L}
\end{array}
\right)
\end{equation}

\noindent are (4x1) column matrices.  We notice here that a matrix
structure is essential for the superspace formulation since the GLA,
in this case, grades an internal symmetry algebra (for example,
Eq.(\ref{oddalgebra2}) involves not just the Hamiltonian, but the internal
symmetry generators as well which would have a matrix representation).
The matrix structure of $Q$ would depend on the internal space upon which
it acts (unlike the usual space-time supersymmetry charges) and the form
given here is appropriate only for the doublet space of $q$ and $\psi$.
Next we introduce the four covariant derivatives as

\begin{eqnarray}
\label{r-covder}
D_{\scriptscriptstyle R} &=& T_-
\frac{\partial}{\partial\theta_{\scriptscriptstyle
R}}+\overline\theta_{\scriptscriptstyle R} T_+ {\cal D}_+\nonumber\\
\overline D_{\scriptscriptstyle R} &=& T_+
\frac{\partial}{\partial\overline\theta_{\scriptscriptstyle
R}}-\theta_{\scriptscriptstyle R} T_- {\cal D}_-
\end{eqnarray}

\begin{eqnarray}
\label{l-covder}
D_{\scriptscriptstyle L} &=& T_-
\frac{\partial}{\partial\theta_{\scriptscriptstyle
L}}+\overline\theta_{\scriptscriptstyle L} T_+ {\cal D}_-\nonumber\\
\overline D_{\scriptscriptstyle L} &=& T_+
\frac{\partial}{\partial\overline\theta_{\scriptscriptstyle
L}}-\theta_{\scriptscriptstyle L} T_- {\cal D}_+
\end{eqnarray}

\noindent and define

\begin{equation}
\label{D-matrix}
{\bf D}=\left(
\begin{array}{cc}
D_{\scriptscriptstyle R} & 0 \\
0 & D_{\scriptscriptstyle L}
\end{array}
\right)
\end{equation}

\begin{equation}
\label{Dbar-matrix}
{\bf\overline D}=\left(
\begin{array}{cc}
\overline D_{\scriptscriptstyle R} & 0 \\
0 & \overline D_{\scriptscriptstyle L}
\end{array}
\right)
\end{equation}

\noindent These covariant derivatives can be easily seen to anticommute
with all supersymmetry charges ${\bf Q}$ and ${\bf \overline Q}$.  In
terms of these covariant derivatives, the Lagrangian of this theory can be
written as

\begin{eqnarray}
\label{matrix-lagrangian}
L &=& {1\over 2}{\displaystyle\sum_{A=R,L}}\int d\theta_A
d\overline\theta_A\left[\left(\overline {\bf D}
\Phi\right)^T\cdot\left({\bf D}\Phi\right)-\Phi^T\cdot\overline {\bf D}
{\bf D}\Phi\right]\nonumber\\
\mbox{}&=& {1\over 2}{\displaystyle\sum_{A=R,L}}\int d\theta_A
d\overline\theta_A\left[\left(\overline
D_A\phi_A\right)^T\cdot\left(D_A\phi_A \right)-\phi_A^T\cdot\overline D_A
D_A\phi_A \right]
\end{eqnarray}

\section{Conclusion}

In this work we have studied a supersymmetric harmonic oscillator
possessing twice as many bosonic variables than fermionic ones.  The model
enjoys a chiral supersymmetry when the fermionic variables are
interchanged with either one of the chiral bosonic sectors.  Besides the
supersymmetries, we have worked out all the global symmetries of the model
and verified that the generators provide a representation of the general
graded Lie algebra GL(2$\vert$1).  We have worked out the physical
spectrum of this model and constructed the superspace formulation.  It is
interesting to see that in the superspace language the separation of the
chiral sectors are clearly displayed, and the charges and the covariant
derivatives carry a matrix structure essentially because the algebra
represents the grading of an internal symmetry group.

\noindent {\bf ACKNOWLEDGEMENTS}  This work has been supported in part by
U.S. Department of Energy, grant No DE-FG-02-91ER 40685, and by CNPq,
Brazilian research agency, Brasilia, Brazil.

\appendix
\renewcommand{\theequation}{\thesection.\arabic{equation}}
\section{Appendix: Dirac Brackets Via Faddeev-Jackiw Formalism}
\setcounter{equation}{0}

First order Lagrangians are constrained systems and must be quantized with
Dirac brackets instead of Poisson brackets\cite{dirac}\cite{hrt}.  The
Dirac brackets of an arbitray first-order system can be calculated with
easy using the technique put forward by Faddeev and Jackiw a few years
ago\cite{fadjack}.  Consider an arbitray system with a finite number of
degrees of freedom $Z_{\scriptscriptstyle A}$, whose Grassman parity is
$\epsilon_{\scriptscriptstyle A}$ and is described by

\begin{equation}
\label{first-order}
L=\dot Z_{\scriptscriptstyle A} K_{\scriptscriptstyle
A}(Z_{\scriptscriptstyle A})- V(Z_{\scriptscriptstyle A})
\end{equation}

\noindent The equations of motion read

\begin{equation}
\label{eq.motion}
\dot Z_{\scriptscriptstyle B} M_{{\scriptscriptstyle B}{\scriptscriptstyle
A}}= - \frac{\partial V}{\partial Z_{\scriptscriptstyle A}}
\end{equation}

\noindent where

\begin{equation}
\label{sympletic}
M_{{\scriptscriptstyle A}{\scriptscriptstyle B}}= \frac{\partial
K^{\scriptscriptstyle B}}{\partial Z_{\scriptscriptstyle A}}-(-1)^{
\epsilon_{\scriptscriptstyle A}\epsilon_{\scriptscriptstyle
B}}\frac{\partial K^{\scriptscriptstyle A}}{\partial Z_{\scriptscriptstyle
B}}
\end{equation}

\noindent is the generalized sympletic matrix\cite{govaerts}.  If the
sympletic matrix is nonsingular, the equation of motion (\ref{eq.motion})
can be solved for the velocities as

\begin{equation}
\label{velocities}
\dot Z_{\scriptscriptstyle A}=(-1)^{ \epsilon_{\scriptscriptstyle
A}}M^{-1}_{{\scriptscriptstyle A}{\scriptscriptstyle B}}\frac{\partial
V}{\partial Z_{\scriptscriptstyle B}}
\end{equation}

\noindent and be written in Hamiltonian form with the introduction of some
generalized or Dirac bracket as

\begin{equation}
\label{dirac-brack}
\left\{Z_{\scriptscriptstyle A}\: , \: Z_{\scriptscriptstyle B}\right\} =
(-1)^{ \epsilon_{\scriptscriptstyle A}}M^{-1}_{{\scriptscriptstyle
A}{\scriptscriptstyle B}}
\end{equation}

\noindent The equations of motion then take the following form

\begin{equation}
\dot Z_{\scriptscriptstyle A}=\left\{Z_{\scriptscriptstyle A}\: , \:
V(Z)\right\}
\end{equation}

\noindent Using (\ref{dirac-brack}) one can verify that the Dirac brackets
for the fermionic variables of the supersymmetric two-dimensional
oscillator are those given in eq.(\ref{ccr2})\cite{edgardo}.


\begin{thebibliography}{30}
\bibitem{susy}For reviews in supersymmetry and supergravity, as well as
references on the original papers see, for instance,  P.Fayet and
S.Ferrara, Phys.Rep.{\bf 32C}(1977)249; J.Wess and J.Bagger,
 ``Supersymmetry and Supergravity'', Princeton University Press, Princeton,
NJ 1983; S.J.Gates, M.T.Grisaru, M.Rocek and W.Siegel, ``Superspace, or
One thousend and One Lessons in Supersymmetry'', Benjamin/Cunnings,
Reading, Mass. 1983,; P.van Nieuwenhuisen, Phys.Rep.{\bf 68C}(1981)264;
M.F.Sohnius, Phys.Rep.{\bf 128C}(1985)40.
\bibitem{gla}P.G.O.Freund and I.Kaplansky, Jour.Math.Phys. {\bf 17}(1976)228.
\bibitem{ssqm}E.Witten, Nucl.Phys.{\bf B188}(1981)513.
\bibitem{ssqm2}P.Salomonson and J.W.van Holten, Nucl.Phys.{\bf B196}(1982)509
and
M.de Combrugghe and V.Rittenberg, Annals of Physics, {\bf 151}(1983)99.
\bibitem{intmod}A.Das and S.Roy, Jour.Math.Phys.{\bf 31}(1990)2145, and
A.Das, W.J.Huang and S.Roy, Intl.Jour.Mod.Phys.{\bf A7}(1992)4293.
\bibitem{das}J.Barcelos-Neto and A.Das, Phys.Rev.{\bf D33}(1986)2863.
\bibitem{dirac}P.A.M.Dirac, Can.J.Math.{\bf 2}(1950)129; Lectures on
Quantum Mechanics (Belfer Graduate School of Science, Yeshiva University,
New York, 1964).
\bibitem{hrt}For a general review, see A. Hanson, T. Regge, and C.
Teitelboim, Constrained Hamiltonian Systems (Academia Nazionale dei
Lincei, Rome, 1976).
\bibitem{fadjack}L.D.Faddeev and R.Jackiw, Phys.Rev.Lett.{\bf 60}(1988)587.
\bibitem{govaerts}J.Govaerts, Intl.J.Mod.Phys.{\bf A5}(1990)3625.
\bibitem{edgardo}J.Barcelos-Neto and E.S.Cheb-Terrab, Z.Phys.{\bf
C54}(1992) 133.
\end{thebibliography}
\end{document}